\documentstyle[prl,preprint,aps]{revtex}
\begin{document}
%\preprint{}
\draft
\title{
\begin{tabbing}
\hspace{12cm} \= {\normalsize UT-ICEPP 94-03}\\
              \> {\normalsize Revised December 1995}
\end{tabbing}            
Local Flux of Low-Energy Antiprotons from 
Evaporating Primordial Black Holes} 
\author{K. Maki\cite{maki}, T. Mitsui, and S. Orito}
\address{Department of Physics, School of Science, 
University of Tokyo, Tokyo 113, Japan} 
%\date{\today}
\maketitle
\begin{abstract}
We investigate low-energy cosmic-ray antiprotons ($\bar{p}$'s) arising from 
the fragmentation of quarks and gluons emitted from evaporating primordial 
black holes (PBHs).  To calculate the local interstellar flux of these 
$\bar{p}$'s, their propagation in the Galaxy is described by a 3-D Monte 
Carlo simulation based on the diffusion model.  This flux is used with 
recent observations to derive new upper limits on (i) the local PBH 
explosion rate ${\cal R}<1.7\times10^{-2}$ pc$^{-3}$yr$^{-1}$, (ii) the 
fraction of the Universe's mass going into PBHs with particular mass, and 
(iii) the average density of PBHs in the Universe.   
\end{abstract} 

%\narrowtext
%--------------
% Introduction
%--------------
{}From a standpoint of cosmological understanding, it would be of great value 
to confirm the existence or nonexistence of primordial black holes (PBHs), 
which may have formed in the early Universe via initial density 
fluctuations, phase transitions, or the collapse of cosmic strings (for a 
review, see Ref.~\cite{C85}).  Hawking \cite{H74} first showed that black 
holes (BHs) emit particles and evaporate by quantum effects, noting that 
PBHs are the only ones with a mass small enough for the quantum emission 
rate to be significant, possibly yielding an observable effect.  For 
example, the hard $\gamma$-rays from small enough PBHs may contribute to the 
diffuse $\gamma$-ray background spectrum, though no distinct signature has 
been observed.  Thus, this leads to an upper limit (U.L.) on the average 
density of PBHs in the Universe, i.e., the ratio of their density to the 
critical density of the Universe, $\Omega_{\rm PBH}$, must be 
$\lesssim10^{-8}$ \cite{PH76}.

Despite such a stringent limit, their signature could still appear in the 
spectrum of cosmic-ray antiprotons ($\bar{p}$'s) \cite{KSWW}, which are 
generally considered to be secondary products from interactions of 
cosmic-ray protons ($p$'s) with the interstellar medium.  This possibility 
arises because, although the kinematics of secondary $\bar{p}$ production by 
such $p$'s should lead to a steep drop in the resultant $\bar{p}$ flux at 
kinetic energies less than 2 GeV, the expected flux of $\bar{p}$'s from PBHs 
(PBH-$\bar{p}$'s) has contrastingly been shown to increase with decreasing 
kinetic energy down to $\sim0.2$ GeV \cite{KSWW}; thus providing a distinct 
signature below 1 GeV.  Hence, searches for such low-energy cosmic-ray 
$\bar{p}$'s could lead to a novel constraint on the density of PBHs, or more 
importantly, demonstrate their existence.  With this in mind, and spurred by 
the recent detection of cosmic-ray $\bar{p}$'s with kinetic energies less 
than 0.5 GeV, being accomplished during a 13-h balloon flight (BESS '93) 
\cite{BESS}, we present a new method describing the propagation of 
$\bar{p}$'s in the Galaxy; thereby obtaining the most accurate-to-date 
spectrum of local interstellar PBH-$\bar{p}$ flux.  We also use these 
results with observed data \cite{BESS} to derive a new U.L. on the density of 
PBHs which are expiring near the Solar system, after which we discuss its 
cosmological aspects assuming their formation via initial density 
fluctuations.    

%-------------------- 
% Theory and Result
%--------------------
The source spectrum of PBH-$\bar{p}$'s is determined using general 
properties of BH evaporation \cite{MW90,M91}.  Briefly, an uncharged, 
non-rotating BH with mass $M$ emits particles with spin $s$ and total 
energy between ($Q$, $Q+dQ$) at a rate \cite{H74}   
\begin{equation}
\frac{dN}{dt}=\frac{\Gamma_{s} dQ}{2\pi\hbar}\left[\exp\left(\frac{Q}
{kT}\right)-(-1)^{2s}\right]^{-1}
\label{eq:HR} 
\end{equation} 
per degree of particle freedom, where $T$ is the BH temperature 
($=\hbar c^{3}/8\pi GMk=1.06\times10^{13}(M/{\rm g})^{-1}$ GeV, with 
$k=1$), and $\Gamma_{s}$ is the dimensionless absorption probability for 
the emitted species.  Considering all species, the corresponding 
mass loss rate can be expressed as \cite{M91} 
\begin{equation} 
\frac{dM}{dt}=-5.34\times10^{25}f(M)\left(\frac{M}{{\rm g}}\right)
^{-2} \;\;\; {\rm g\,\,s}^{-1}, 
\label{eq:Mdot} 
\end{equation} 
where $f(M)$, a function of the number of emitted species, is normalized 
to unity for large $M$ ($\gg10^{17}$ g) and increases with decreasing 
$M$.  From Eq.~(\ref{eq:Mdot}), it follows that 
$M_{*}\simeq5.3\times10^{14}(t_{u}/16\;{\rm Gyr})^{1/3}$ g, or 
$3\times10^{-19}(t_{u}/16\;{\rm Gyr})^{1/3}$ in units of solar mass, 
where $M_{*}$ is the initial mass of a PBH expiring today; i.e., its initial 
lifetime equals the present age of the Universe $t_{u}$, being taken here 
as 16 Gyr \cite{KT90}.  PBHs with initial mass $M_{i}<M_{*}$ should have 
completely evaporated by now, while those with $M_{i}$ slightly larger 
than $M_{*}$ have an extremely high present temperature and will soon 
expire by ``explosion.'' 
 
Under the assumptions that (i) Eq.~(\ref{eq:HR}) holds for each emitted 
species of quarks and gluons and (ii) all PBH-$\bar{p}$'s are their 
fragments, which is consistent with observed $e^{+}e^{-}$ annihilations, 
we calculate the PBH-$\bar{p}$ source spectrum per unit volume via the 
following three-step procedure: (1) The JETSET 7.4 \cite{LUND} Monte 
Carlo simulation code is used to obtain the fragmentation function 
$dg^{(j)}_{\bar{p}}(E_{\bar{p}}, Q)/dE_{\bar{p}}$ describing the 
fragmentation of each emitted species $j$ with total energy $Q$ into 
$\bar{p}$'s with total energy $E_{\bar{p}}$; (2) The $\bar{p}$ emission 
spectrum from a BH at present temperature $T$, 
$d\Phi_{\bar{p}}(E_{\bar{p}}, T)/dE_{\bar{p}}$, is calculated by 
convolving Eq.~(\ref{eq:HR}) with 
$dg^{(j)}_{\bar{p}}(E_{\bar{p}}, Q)/dE_{\bar{p}}$ for $Q\geq E_{\bar{p}}$ 
and summing over all $j$ and their degrees of freedom; and (3) The 
expected PBH-$\bar{p}$ source spectrum per unit volume is obtained by 
convolving the PBH present temperature distribution $dn/dT$ with 
$d\Phi_{\bar{p}}(E_{\bar{p}}, T)/dE_{\bar{p}}$ for $T\gtrsim0.1$ GeV, 
where $n$ is the number of PBHs per unit volume.  Under 
Eq.~(\ref{eq:Mdot}), $dn/dT$ versus $T$ ($\gtrsim0.1$ GeV) is roughly a 
power-law function ($\propto T^{-4.2}$), with its normalization being 
solely determined by the value of the initial mass spectrum $dn/dM_{i}$ 
at $M_{i}=M_{*}$.  

We checked the reliability of our use of JETSET 7.4 \cite{LUND} via data 
from $e^{+}e^{-}$ colliders, i.e., observed $\bar{p}$ spectra obtained 
from the fragmentation of quarks and gluons at various $Q$ relevant to 
evaporating PBHs.  First, we verified that the predominant contribution 
to PBH-$\bar{p}$'s occurs at $Q=1$--5 GeV, after which the simulated 
$\bar{p}$ spectra from quark fragmentation were compared with DASP data at 
$Q=1.8$--2.5 GeV \cite{DATA} and ARGUS data at $Q=4.99$ GeV \cite{DATA}, 
while those from gluon fragmentation with ARGUS data from the direct 
decay of $\Upsilon(1S)$, which mainly proceeds through three gluons with 
an average energy of $m_{\Upsilon(1S)}/3=3.2$ GeV \cite{DATA}.  Resultant 
comparisons showed good agreement (data not shown).  

Since the propagation of PBH-$\bar{p}$'s in the Galaxy has thus far only 
been roughly treated \cite{KSWW}, we calculate the local interstellar 
PBH-$\bar{p}$ flux from the PBH-$\bar{p}$ source spectrum utilizing a 3-D 
Monte Carlo simulation code based on the diffusion model; an expansion of 
the 1-D simulation by Owens and Jokipii \cite{OJ}.  We assume that the PBH 
spatial distribution is proportional to the mass density distribution 
within the Galactic halo, i.e., $\propto(1+(r/r_{c})^{2})^{-1}$ 
\cite{iso}, where $r$ is a distance from the Galactic center and $r_{c}$ 
is the core radius of 7.8 kpc.  As shown later, the local PBH-$\bar{p}$ 
flux arises only from nearly expired PBHs existing within a few kpc away 
from the Solar system, whose location is at $r=r_{\odot}=8.5$ kpc.  
Accordingly, the local PBH-$\bar{p}$ flux can be calculated via the 
three-step procedure by simply using 
$dn(M_{i}=M_{*}, r=r_{\odot})/dM_{i}$, which is parametrized by 
introducing an unknown parameter $\varepsilon_{*}$:   
\begin{equation}
\frac{dn}{dM_{i}}(M_{i}=M_{*}, r=r_{\odot})
=\varepsilon_{*}\frac{\rho_{h\odot}}{M_{*}^{2}}, 
\label{eq:estar}
\end{equation}
where $\rho_{h\odot}$ is the local density of halo dark matter 
($\simeq0.3$ GeV~cm$^{-3}$ \cite{darksolar}).  Under 
Eq.~(\ref{eq:estar}), $\varepsilon_{*}$ represents the ratio of the 
density of PBHs with $M_{i}=M_{*}$ to $\rho_{h\odot}$.  

Details regarding the diffusive propagation of $\bar{p}$'s will be described 
elsewhere \cite{mitsui5yr}.  Briefly, however, each $\bar{p}$ ejected from 
the fragmentation of quarks and gluons is assigned to an initial position 
${\bf x}_0$ and energy $E_0$, after which it travels ${\bf x}$ to 
${\bf x}+{\bf u}\sqrt{6D\Delta t}+(\mbox{\boldmath $\nabla$}D)\Delta t$ in 
each subsequent time step $\Delta t$, where ${\bf u} \sqrt{6D\Delta t}$
represents the effect of isotropic diffusion during $\Delta t$, ${\bf u}$ is 
a unit vector with random direction, $D = D({\bf x})$ is the diffusion 
coefficient, and $(\mbox{\boldmath $\nabla$}D)\Delta t$ expresses the 
anisotropy of diffusion caused by the spatial gradient of $D$.  We do not 
consider the effect of convection due to galactic wind, because Webber 
{\it et al.}\ \cite{Web} showed that it does not significantly affect 
cosmic-ray propagation.  Regarding subsequent collisions and energy loss,
the interstellar medium (ISM) is considered to consist of 90\% hydrogen and 
10\% helium atoms \cite{Eng}, and its number density distribution is modeled 
as $n_{atom}=1.1\exp(-z/100\,{\rm pc})$ atoms/cm$^3$, where $z$ is the 
perpendicular distance from the galactic plane ($z = 0$).  While propagating 
through the ISM, $\bar{p}$'s lose energy by ionization and are lost by 
annihilation.  Those $\bar{p}$'s passing near the Solar system 
($\lesssim$ 25 pc) are included in the flux calculation.

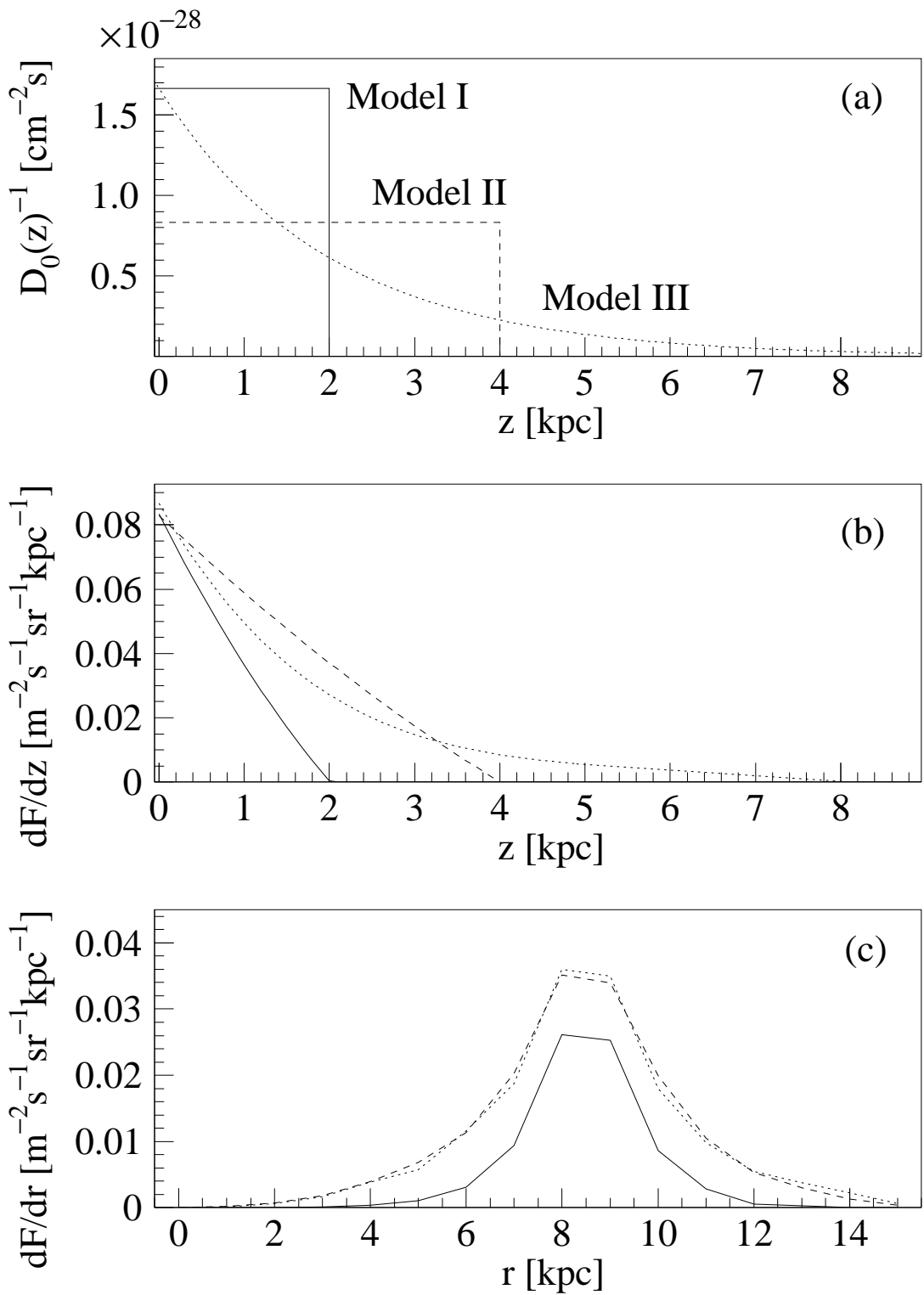
\begin{figure}
\caption{(a)~Distribution of $D_0(z)^{-1}$ in halo models I--III as shown 
in the $z$-direction where the Solar system is located on the galactic 
plane ($z=0$).  (b)~Distribution of PBHs contributing to the local 
$\bar{p}$ flux $F$ near the Solar system, obtained by assuming 
$\varepsilon_{*}=1.0\times10^{-8}$, as shown in the $z$-direction.  
(c)~The same as (b) shown in the $r$-direction where the Solar system 
is located at $r=8.5$ kpc.}
\label{fig1}
\end{figure}

The Galaxy is modeled as a cylindrical diffusing halo \cite{Web} with a 
diameter of 40 kpc and halo thickness of 2$h$.  Free escape is assumed to 
occur at the boundaries.  As normally done, we parametrize the diffusion 
coefficient as $D = D_0(z) D_1(R)$, where $R$ is the rigidity of the particle.
We apply three halo models for $D_0(z)$ (Fig.~\ref{fig1}(a) shows 
$D_0(z)^{-1}$): two with constant $D_0$ but different values of $h$ 
(models I and II), and one in which $D_0$ is dependent on $z$ (model III).  
Assuming $D_0$ is constant, $D_0/h$ and $D_1(R)$ are determined \cite{Web} as 
$(8\pm 1.6)\times 10^5$ cm~s$^{-1}$ and $\sim (R/{\rm GV})^{0.6}$, 
respectively, which fits the secondary to primary ratios of cosmic-ray 
nuclei, e.g., the ratios of boron to carbon (B/C) and sub-iron to iron 
(sub-Fe/Fe).  The value of $h$ is also constrained from 2 to 4 kpc \cite{Web} 
in order to fit the radioactive secondary to stable secondary ratios, e.g., 
the ratios of radioactive beryllium to stable beryllium ($^{10}$Be/$^{9}$Be).  
Thus, $h=2$ and 4 kpc in models I and II, respectively.  Model III also fits 
the above ratios at the same $D_1(R)$.

The simulated distribution of PBHs contributing to the integrated $\bar{p}$ 
flux $F$ near the Solar system, obtained by assuming 
$\varepsilon_{*}=1.0\times10^{-8}$, is shown for halo models I--III in 
Fig.~\ref{fig1}(b) and (c) in the $z$- and $r$-direction, respectively.  
Note that all models show the same contribution at $z=0$ (galactic plane) as 
they are constrained to reproduce the observed secondary to primary ratios 
which originate from secondary production occurring there.  In addition, 
only PBHs within a few kpc away from the Solar system contribute a 
substantial flux.  When considering these results along with the fact that 
the mass density within this region is relatively constrained 
\cite{darksolar}, this indicates that the Solar system $\bar{p}$ flux is only 
slightly dependent on the assumed mass density distribution within the 
Galactic halo \cite{iso}.  The $\bar{p}$ mean confinement time is also 
calculated by the code as 4.0, 7.9, and 8.0 $\times 10^7$ yr for models 
I--III, resulting in $F=0.78$, 1.53, and 
$1.55\times10^{-1}$ m$^{-2}$s$^{-1}$sr$^{-1}$, respectively.

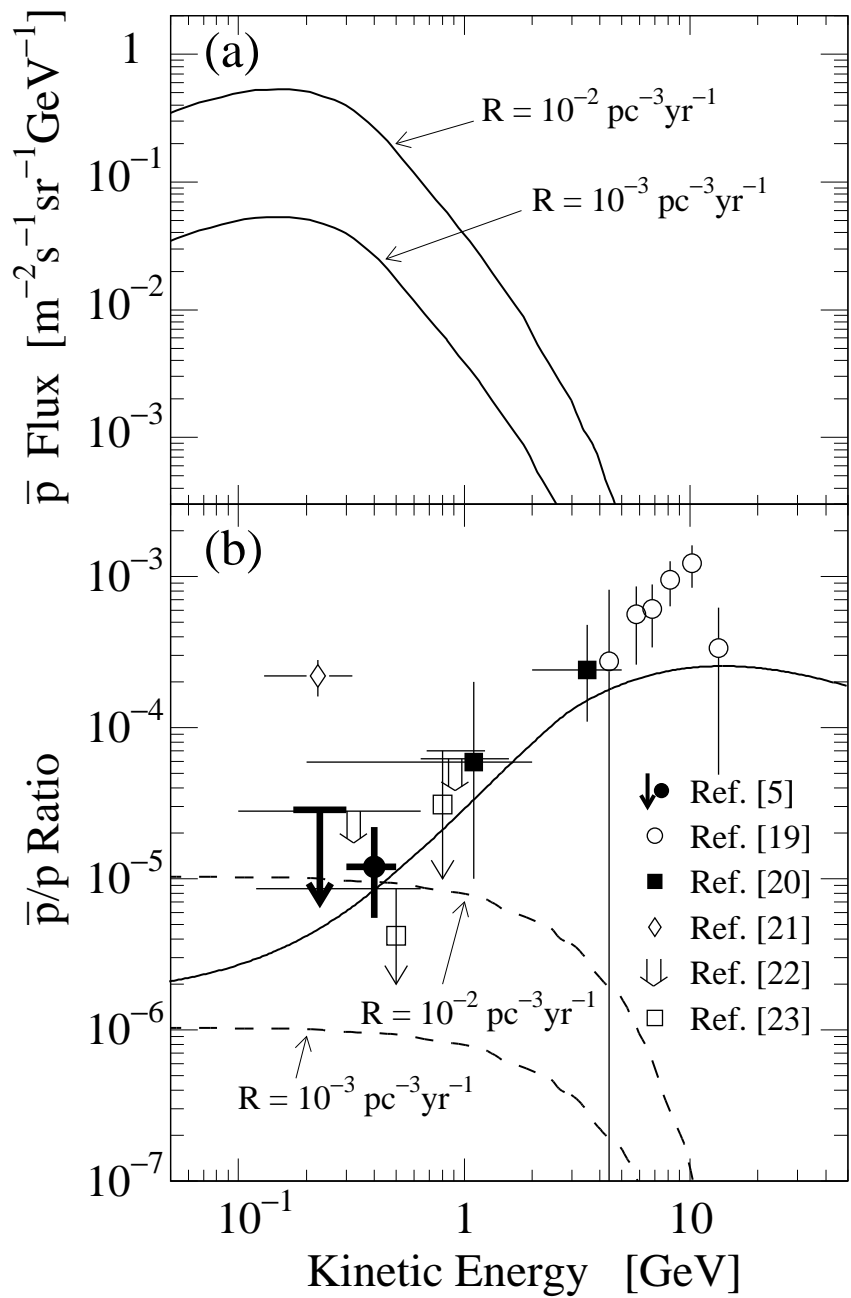
\begin{figure}
\caption{(a)~Simulated local interstellar PBH-$\bar{p}$ flux using model 
III with ${\cal R}=1\times10^{-2}$ and $1\times10^{-3}$ pc$^{-3}$yr$^{-1}$.  
(b)~Corresponding $\bar{p}/p$ ratio for PBH-$\bar{p}$'s at the top of the 
atmosphere (dashed lines), the expected $\bar{p}/p$ ratio for secondary 
$\bar{p}$'s (solid line) \protect\cite{mitsui5yr}, and observational data 
\protect\cite{BESS,G84,Bo87,BSP81,PBAR,LEAP}.}
\label{fig2}
\end{figure}

As we have shown that the local PBH-$\bar{p}$ flux can only be due to 
contributions from PBHs that are close to explosion and exist within a 
few kpc away from the Solar system, data from searches in which 
low-energy cosmic-ray $\bar{p}$'s are detected can {\em directly} 
constrain (or possibly reveal) the PBH explosion rate averaged over this 
local region, termed here as the local PBH explosion rate ${\cal R}$, i.e.,  
\begin{eqnarray}
{\cal R} & \equiv & \frac{dn}{d\tau_{i}}(\tau_{i}=t_{u}, r=r_{\odot}) 
\nonumber \\ 
 & = & \frac{dn}{dM_{i}}(M_{i}=M_{*}, r=r_{\odot})\times
\frac{M_{*}}{3t_{u}}, 
\label{eq:R}
\end{eqnarray}
where $\tau_{i}$ ($=t_{u}(M_{i}/M_{*})^{3}$ at $M_{i}\approx M_{*}$) is 
initial lifetime of PBHs with initial mass $M_{i}$.  Figure~\ref{fig2}(a) 
shows simulated local interstellar PBH-$\bar{p}$ flux using model III 
with two possible values of ${\cal R}$, i.e., $1\times10^{-2}$ and 
$1\times10^{-3}$ pc$^{-3}$yr$^{-1}$.  Then, to compare this $\bar{p}$ 
flux with observational data, we converted it into the $\bar{p}/p$ ratio 
at the top of the atmosphere (TOA) by assuming that (i) the interstellar 
$p$ flux $J_{p}=1.5\times10^{4}\beta^{-1}(R/{\rm GV})^{-2.74}$ 
m$^{-2}$s$^{-1}$sr$^{-1}$GeV$^{-1}$ \cite{IS-P}, where $\beta$ is the 
ratio of particle velocity to the velocity of light, and (ii) the 
corresponding flux of both $p$'s and $\bar{p}$'s at TOA can be similarly 
calculated by accounting for solar modulation using the force-field 
approximation \cite{GA68} with modulation parameter $\phi=650$ MV, which 
corresponds to that obtained from the BESS '93 flight \cite{BESS}.  
Figure~\ref{fig2}(b) compares the resultant $\bar{p}/p$ ratio for 
PBH-$\bar{p}$'s with the expected $\bar{p}/p$ ratio for secondary 
$\bar{p}$'s \cite{mitsui5yr} and observational data 
\cite{BESS,G84,Bo87,BSP81,PBAR,LEAP}.  Note that no distinct signature of 
evaporating PBHs is apparent, i.e., at kinetic energies below 1 GeV, the 
data shows no tendency to reach a constant $\bar{p}/p$ ratio.  As no 
signature exists, statistical analysis of recent observations \cite{BESS} 
leads to the following U.L. on ${\cal R}$ with 90\% confidence level (C.L.): 
\begin{equation}
{\cal R}<1.7\times10^{-2}\;\;{\rm pc}^{-3}{\rm  yr}^{-1},
\label{eq:Rlim}
\end{equation}
which is almost eight orders of magnitude more stringent than the present 
U.L. on the rate of 50-TeV $\gamma$-ray bursts 
(${\cal R}<8.5\times10^{5}$ pc$^{-3}$yr$^{-1}$ \cite{CYG}), and the 
practical sensitivity for 100-MeV $\gamma$-ray bursts 
(${\cal R}\sim10^{6}$ pc$^{-3}$yr$^{-1}$ \cite{S94}).  Note that the 
observed diffuse $\gamma$-ray background spectrum, which contains 
contributions from all evaporating PBHs throughout the Universe, does not 
directly constrain ${\cal R}$.
 
%------------ 
% Discussion
%------------
Equations~(\ref{eq:estar})--(\ref{eq:Rlim}) give the following U.L. 
on $\varepsilon_{*}$ with 90\% C.L.:
\begin{equation}
\varepsilon_{*}<2.8\times10^{-8}\left(\frac{\rho_{h\odot}}{0.3\;{\rm GeV}
\,{\rm cm}^{-3}}\right)^{-1}
\left(\frac{t_{u}}{16\;{\rm Gyr}}\right)^{4/3},
\label{eq:elim} 
\end{equation} 
which can be used to derive an U.L. on the fraction of the Universe's 
mass going into PBHs with mass $M_{*}$, i.e., $\beta(M_{*})$ \cite{C85}.  
If PBHs are assumed to have formed via initial density fluctuations, 
$\beta(M_{*})$ should be closely related to their amplitude on a scale of 
$M_{*}$.  Thus, under this assumption, and for a flat Friedmann universe, 
Eq.~(\ref{eq:estar}) gives 
$\beta(M_{*})\sim1\times10^{-18}\varepsilon_{*}(\Omega_{h}/0.1)$, where 
$\Omega_{h}$ is the ratio of the density of halo dark matter to the 
critical density of the Universe.  Finally, assuming $\Omega_{h}=0.1$ 
\cite{KT90}, Eq.~(\ref{eq:elim}) leads to an U.L. of 
$\beta(M_{*})<3\times10^{-26}$.  

Further, by assuming that such initial density fluctuations are 
scale-invariant, the initial mass spectrum of PBHs should have a 
power-law form, i.e., $dn/dM_{i}\propto M_{i}^{-\alpha}$ \cite{C85}, 
where $\alpha=5/2$ if the Universe was radiation-dominated when PBHs 
formed.  Using this initial mass spectrum normalized by 
Eq.~(\ref{eq:estar}) at $M_{i}=M_{*}$, we integrate 
$M_{i}\times dn/dM_{i}$ over $M_{*}\leq M_{i}\leq\infty$ to obtain 
$\Omega_{\rm PBH}\equiv\varepsilon_{*}\Omega_{h}\int_{1}^{\infty}
x^{1-\alpha}dx=2\times10^{-1}\varepsilon_{*}(\Omega_{h}/0.1)$, where 
$x\equiv M_{i}/M_{*}$.  Finally, assuming $\Omega_{h}=0.1$ \cite{KT90}, 
Eq.~(\ref{eq:elim}) leads to an U.L. of 
$\Omega_{\rm PBH}<6\times10^{-9}$, being comparable to that from the 
diffuse $\gamma$-ray background spectrum ($\lesssim10^{-8}$) \cite{PH76}.

%------------
% Conclusion
%------------
In closing, if future long-duration ($\sim8$ days) balloon flights allow 
us to precisely measure the cosmic-ray $\bar{p}$ flux at kinetic energies 
from 0.2 to 2 GeV, the U.L. on ${\cal R}$ can be significantly reduced to 
$\sim2\times10^{-3}$ pc$^{-3}$yr$^{-1}$, or more importantly, such 
observations could confirm the existence of evaporating PBHs.  
  
%-----------------  
% Acknowledgement
%-----------------
Sincere gratitude is extended to all members collaborating on the 
BESS experiment.  K.~M. acknowledges a fellowship from the Japan 
Society for the Promotion of Science.  

%------------
% References 
%------------

%-----------------
% figure captions 
%-----------------

\begin{references}
\bibitem[*]{maki} 
Electronic address: maki@icepp.s.u-tokyo.ac.jp
\bibitem{C85} 
B.~J.~Carr, in {\it Observational and Theoretical 
Aspects of Relativistic Astrophysics and Cosmology}, edited by J.~L.~Sanz 
and L.~J.~Goicoechea (World Scientific, Singapore, 1985), p.~1.
\bibitem{H74} 
S.~W.~Hawking, Commun. Math. Phys. {\bf 43}, 199 (1975).
\bibitem{PH76} 
D.~N.~Page and S.~W.~Hawking, Astrophys. J. {\bf 206}, 1 (1976).
\bibitem{KSWW} 
P.~Kiraly {\it et al.}, Nature (London) {\bf 293}, 120 (1981); 
M.~S.~Turner, {\it ibid.} {\bf 297}, 379 (1982); 
J.~H.~MacGibbon and B.~J.~Carr, Astrophys. J. {\bf 371}, 447 (1991). 
\bibitem{BESS}
K.~Yoshimura {\it et al.}, Phys. Rev. Lett. {\bf 75}, 3792 (1995).  
\bibitem{MW90} 
J.~H.~MacGibbon and B.~R.~Webber, Phys. Rev. D {\bf 41}, 3052 (1990).
\bibitem{M91}
J.~H.~MacGibbon, Phys. Rev. D {\bf 44}, 376 (1991).
\bibitem{KT90}
E.~W.~Kolb and M.~S.~Turner, {\it The Early Universe} (Addison-Wesley, 
New York, 1990).  
\bibitem{LUND} 
T.~Sj\"{o}strand, Comput. Phys. Commun. {\bf 82}, 74 (1994).
\bibitem{DATA} 
DASP Collab., R.~Brandelik {\it et al.}, Nucl. Phys. {\bf B148}, 189 
(1979); 
ARGUS Collab., H.~Albrecht {\it et al.}, Z. Phys. C {\bf 44}, 547 (1989). 
\bibitem {OJ} 
A.~J.~Owens and J.~R.~Jokipii, Astrophys. J. {\bf 215}, 677 (1977); 
{\bf 215}, 685 (1977).
\bibitem {iso} 
J.~A.~R.~Caldwell and J.~P.~Ostriker, Astrophys. J. {\bf 251}, 61 (1981).
\bibitem {darksolar} 
R.~A.~Flores, Phys. Lett. {\bf B215}, 73 (1988).
\bibitem {mitsui5yr} 
T.~Mitsui, in preparation.
\bibitem {Web} 
W.~R.~Webber, M.~A.~Lee, and M.~Gupta, Astrophys. J. {\bf 390}, 96 (1992).
\bibitem {Eng} 
J.~J.~Engelmann {\it et al.}, Astron. Astrophys. {\bf 233}, 96 (1990).
\bibitem{IS-P} 
E.~S.~Seo {\it et al.}, Astrophys. J. {\bf 378}, 763 (1991).  
\bibitem{GA68} 
L.~J.~Gleeson and W.~I.~Axford, Astrophys. J. {\bf 154}, 1011 (1968).
\bibitem{G84}  
R.~L.~Golden {\it et al.}, Phys. Rev. Lett. {\bf 43}, 1196 (1979); 
Astrophys. Lett. {\bf 24}, 75 (1984). 
\bibitem{Bo87} 
E.~A.~Bogomolov {\it et al.}, Proc. 20th Int. Cosmic Ray Conf. (Moscow) 
{\bf 2}, 72 (1987); 
Proc. 21st Int. Cosmic Ray Conf. (Adelaide) {\bf 3}, 288 (1990).
\bibitem{BSP81} 
A.~Buffington, S.~M.~Schindler, and C.~R.~Pennypacker, Astrophys. J. 
{\bf 248}, 1179 (1981). 
\bibitem{PBAR} 
S.~P.~Ahlen {\it et al.}, Phys. Rev. Lett. {\bf 61}, 145 (1988);  
M.~H.~Salamon {\it et al.}, Astrophys. J. {\bf 349}, 78 (1990).
\bibitem{LEAP} 
R.~E.~Streitmatter {\it et al.}, Proc. 21st Int. Cosmic Ray Conf. 
(Adelaide) {\bf 3}, 277 (1990);  
A.~Moats {\it et al.}, {\it ibid.} {\bf 3}, 284 (1990);  
S.~J.~Stochaj, Ph.D Thesis, Univ. of Maryland (1990).  
\bibitem{CYG} 
D.~E.~Alexandreas {\it et al.}, Phys. Rev. Lett. {\bf 71}, 2524 (1993).
\bibitem{S94} 
D.~V.~Semikov, Astrophys. J. {\bf 436}, 254 (1994).
\end{references}
\end{document}